\pdfoutput=1  
\documentclass[]{article}  
\usepackage{url,float}
\usepackage{graphicx}
\usepackage{amsfonts}
\usepackage{amssymb}
\usepackage{latexsym}

\newcommand{\hide}[1]{}

\newcommand{\ABox}{
\raisebox{3pt}{\framebox[6pt]{\rule{6pt}{0pt}}}
}
\newenvironment{proof}{{\bf Proof:}}{\hfill\ABox}

\newtheorem{theorem}{{\bf Theorem}}

\newtheorem{lemma}[theorem]{Lemma}

\newcommand{\lemlab}[1]{\label{lemma:#1}}

\newcommand{\figlab}[1]{\label{fig:#1}}
\newcommand{\seclab}[1]{\label{sec:#1}}

\newcommand{\eqref}[1]{\ref{eq:#1}}
\newcommand{\figref}[1]{\ref{fig:#1}}

{\makeatletter
 \gdef\xxxmark{%
   \expandafter\ifx\csname @mpargs\endcsname\relax 
     \expandafter\ifx\csname @captype\endcsname\relax 
       \marginpar{xxx}
     \else
       xxx 
     \fi
   \else
     xxx 
   \fi}
 \gdef\xxx{\@ifnextchar[\xxx@lab\xxx@nolab}
 \long\gdef\xxx@lab[#1]#2{{\bf [\xxxmark #2 ---{\sc #1}]}}
 \long\gdef\xxx@nolab#1{{\bf [\xxxmark #1]}}
 \gdef\turnoffxxx{\long\gdef\xxx@lab[##1]##2{}\long\gdef\xxx@nolab##1{}}%
}

\def\d{{\delta}}
\def\e{{\epsilon}}

\newcommand{\squeezelist}{\setlength{\itemsep}{0pt}}

\title{Band Unfoldings and Prismatoids:\\A Counterexample}

\author{%
Joseph O'Rourke%
    \thanks{Dept. Comput. Sci., Smith College, Northampton, MA
      01063, USA.
      \protect\url{orourke@cs.smith.edu}.}
}

\begin{document}
\maketitle

\begin{abstract}
This note shows that the hope expressed in~\cite{adlmost-eunpb-07}---that
the new algorithm for edge-unfolding any polyhedral band without overlap
might lead to an algorithm for unfolding any prismatoid without overlap---cannot
be realized.  A prismatoid is constructed whose sides constitute a nested
polyhedral band, with the property that every placement of the prismatoid
top face overlaps with the band unfolding.
\end{abstract}

\section{Introduction}
\seclab{intro}
An \emph{edge unfolding} of a polyhedron is a cutting of its surface
along edges so that it unfolds flat to a single, non-overlapping piece
in the plane.  Few classes of convex polyhedra are known to have
edge unfoldings.  In particular, it is unknown if the
class of prismatoids have edge unfoldings.
A \emph{prismatoid} is the convex hull of two convex polygons $A$ and $B$ lying
in parallel planes.
See~\cite[Chap.~22]{do-gfalop-07} for background on this problem.

Recently an edge unfolding result has been obtained for bands.
A {\em polyhedral band} is the intersection of the surface of a convex
polyhedron with the space between two parallel planes, as long as this
space does not contain any vertices of the polyhedron.
In~\cite{adlmost-upb-05}~\cite{adlmost-eunpb-07},
it was established that \emph{nested polyhedral bands},
where the top polygon projects orthogonally inside the bottom or vice versa,
have an edge unfolding.
This result was extended in~\cite{a-rps-05} to remove the nesting
restriction, and to permit vertices along the top and bottom rims.

A prismatoid can be viewed as a polyhedral band closed by top
and bottom convex polygons $A$ and $B$.
As there has been no success achieving a ``volcano unfolding''
of a prismatoid~\cite[pp.~321--3]{do-gfalop-07},
the suggestion made in~\cite{adlmost-eunpb-07} is natural:

\begin{quotation}
\noindent
``with it established that arbitrary bands can be unfolded without overlap,
it remains interesting to see whether this can lead to a non-overlapping
unfolding of prismatoids, including the top and bottom faces.
It is natural to hope that these faces could be nestled on opposite sides
of the unfolded band, but we do not know how to ensure non-overlap.''
\end{quotation}

The purpose of this note is to dash this hope by constructing a
prismatoid, whose side faces constitute a nested polyhedral band,
such that, for every edge unfolding of the band, the top $A$ cannot be
attached to the band without overlap.

\section{Prismatoid Description}
There are two key aspects to the design of the prismatoid counterexample.
First, the top face $A$ has three acute angles.  It is a hexagon
formed by replacing each side of an equilateral triangle with two
nearly collinear edges.
Figure~\figref{prismatoids_3D}(a) illustrates the basic structure,
exaggerating the 3D height.
Second, the prismatoid is nearly flat, 
as in~(b) of the figure.

\begin{figure}[htbp]
\centering
\includegraphics[width=\linewidth]{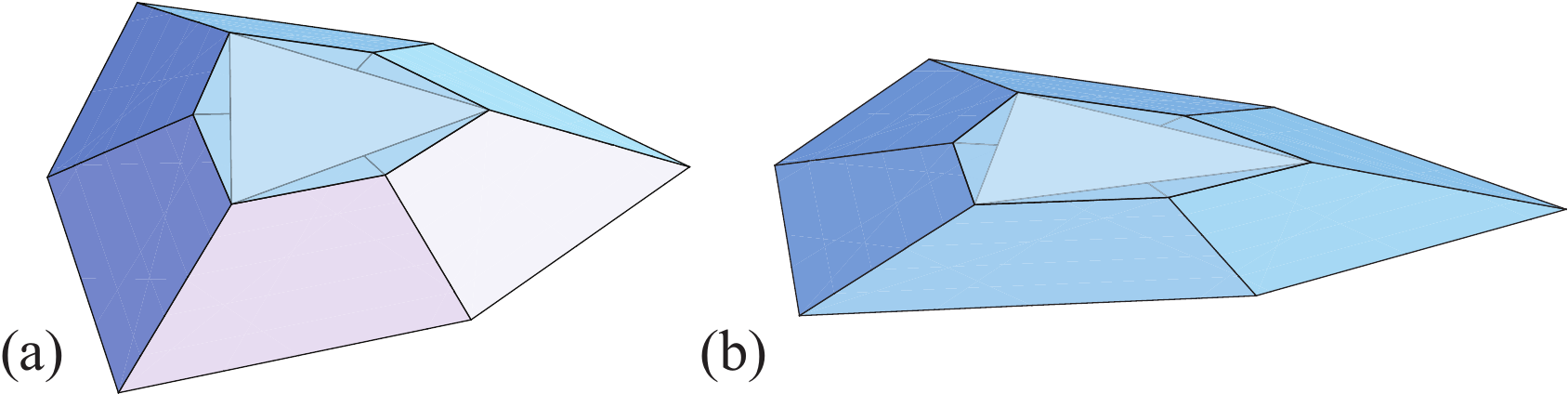}
\caption{(a)~The prismatoid with $z/y = 1/2$;
(b)~The prismatoid with $z/y= 0.19$ leading to $2\d=\e \approx 1^\circ$.}
\figlab{prismatoids_3D}
\end{figure}

Figure~\figref{flat_prismatoid} shows an overhead view
of the prismatoid, and establishes
notation.
The six vertices of $A$ are $(a_0,\ldots,a_5)$,
and each is connected to its counterpart $b_i$ on the bottom face $B$.
The side faces $(a_i,b_i,b_{i+1},a_{i+1})$ are each parallelograms.
Three parameters determine the dimensions of the shape:
\begin{enumerate}
\squeezelist
\item $z$, the distance between the planes containing $A$ and $B$.
\item $y$, the $y$-distance between $a_0$ and $b_0$, as illustrated in the figure.
\item $h$, the offset of $a_0$ with respect to the edge of the equilateral
triangle inscribed in $A$ (again illustrated in the figure).
\end{enumerate}
Let $\d$ be the curvature, i.e., the angle deficit, at $a_0$, and
let $\e$ be the curvature at $a_1$.  Cutting and flattening a vertex opens
it by an amount equal to the curvature.
By choosing $z/y$ small, we can make these curvatures small.
And---although this is not essential---adjustment of the parameters
permits achieving $\d = \frac{1}{2}\e$
(e.g., as in Figure~\figref{prismatoids_3D}(b)).

The counterexample works for a variety of combinations of the
three parameters.
No attempt has been made to delimit the precise set of parameter
combinations that lead to overlap,
as choosing $\d$ and $\e$ sufficiently small suffices.

\begin{figure}[htbp]
\centering
\includegraphics[width=0.75\linewidth]{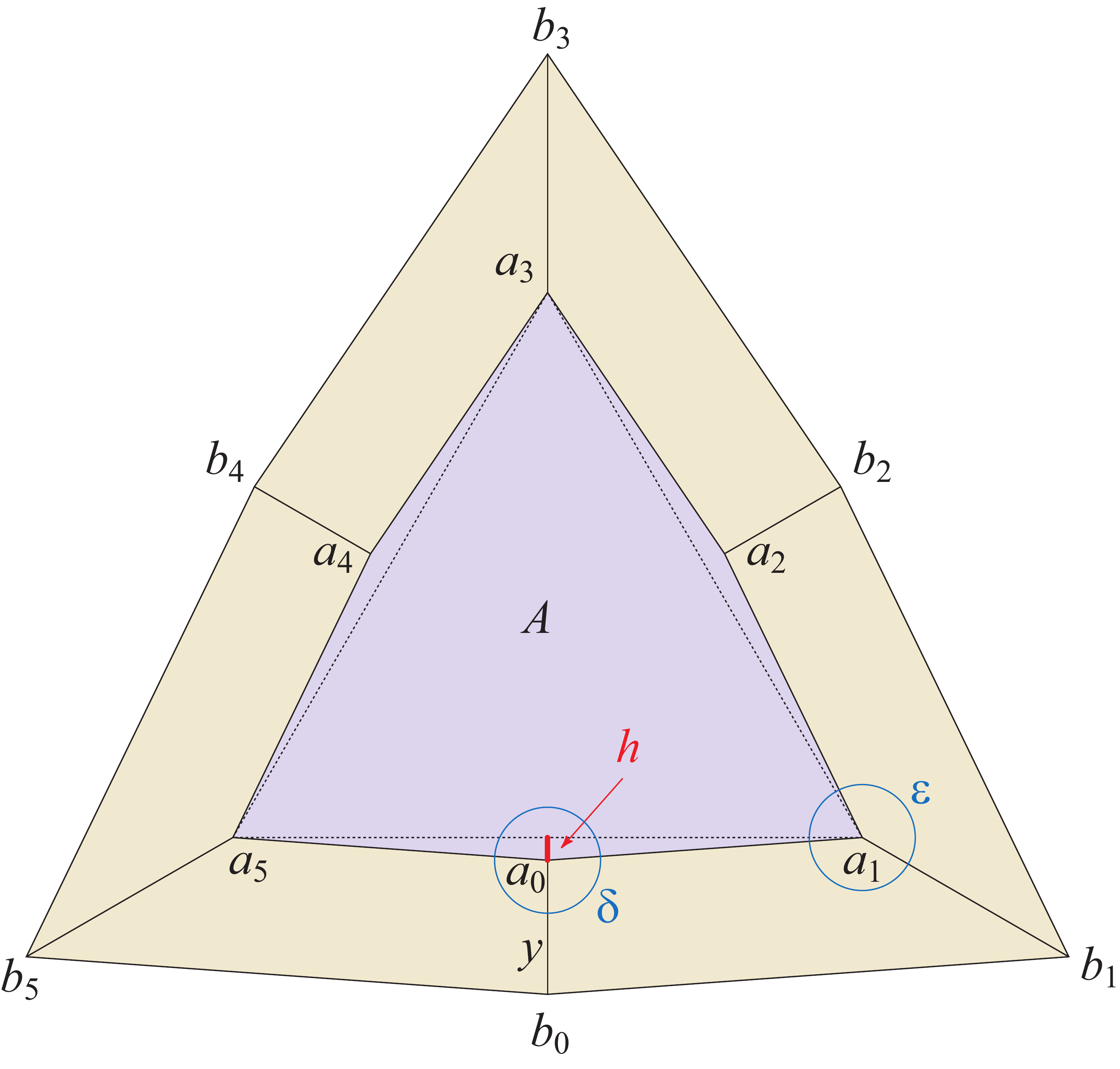}
\caption{Overhead view of prismatoid.
If an origin is set at the midpoint of the bottom side of the equilateral
triangle on $A$, then
$a_0=(0,-h,0)$ and $b_0 = (0,-(h+y),-z)$.}
\figlab{flat_prismatoid}
\end{figure}

\section{Unfolding}
Because of the symmetry of $A$, there are only two distinct band edge-cuts
that need be considered:  cutting at the apex of the equilateral triangle,
say, edge $a_3 b_3$, or cutting at a side vertex, say, edge $a_0 b_0$.
Because we choose $\d$ and $\e$ to be small, the band opens only a little.
Figure~\figref{hexa_cex}(a-c) illustrates the opening at $a_3$,
and (d-f) the opening at $a_0$.
For the opening at $a_3$, there are only three distinct edges at which to
attach $A$: along edge $a_0 a_1$, or $a_1 a_2$, or $a_2 a_3$.
All three attachments lead to overlaps.
Similarly, cutting at $a_0$ leaves three distinct attachment
edges, $a_3 a_4$, $a_4 a_5$, and $a_5 a_0$, each of which leads to overlap.
\begin{figure}[htbp]
\centering
\includegraphics[width=\linewidth]{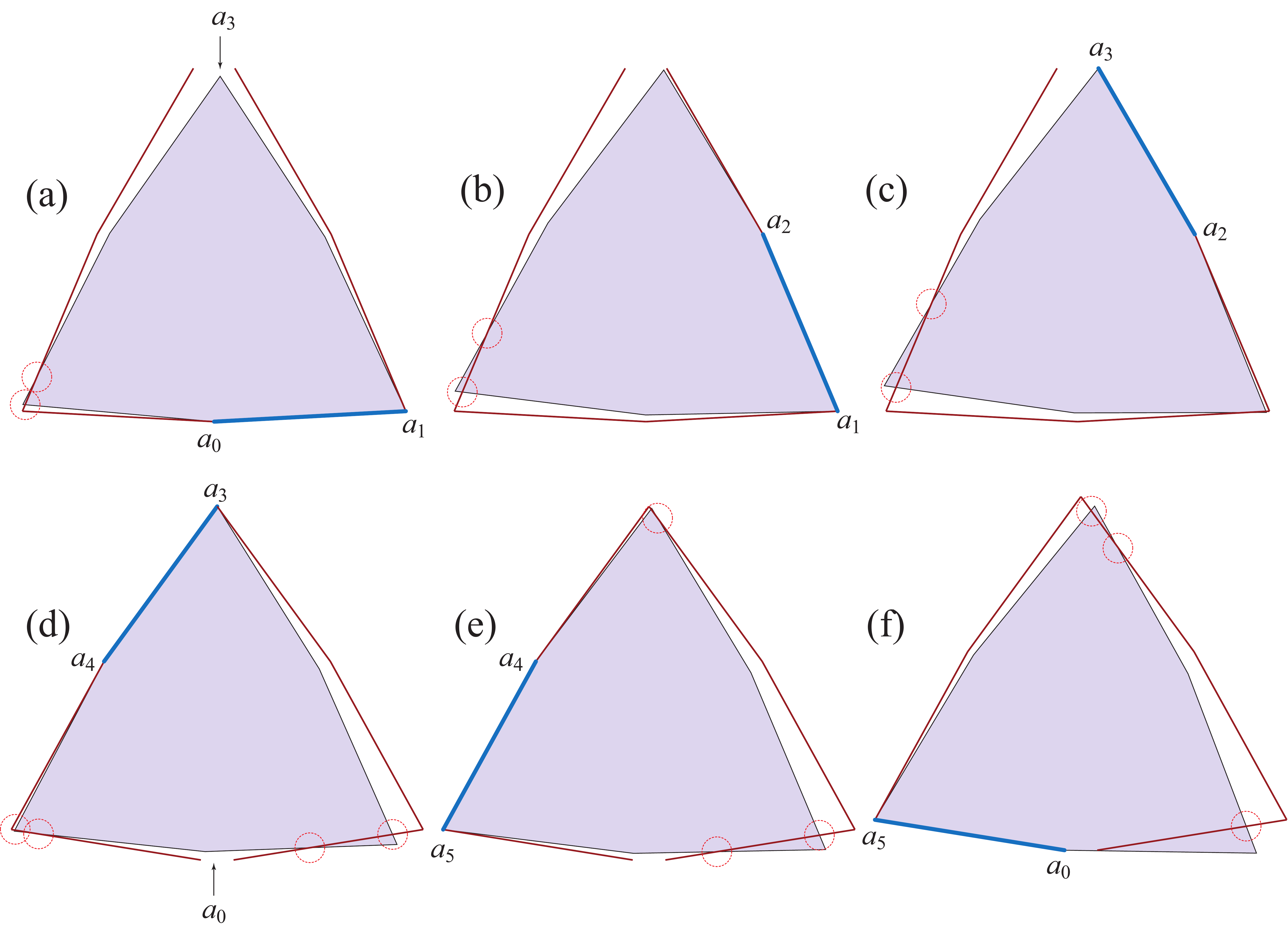}
\caption{Placements of $A$ when $a_3$ is cut
(top row) and when $a_0$ is cut (bottom row).
The band rim is shown red; the band itself lies
outside the rim (cf.~Fig.~\protect\figref{prismatoids_3D}(b)).
The attachment edge of the band to $A$ is blue.
Circles indicate overlap.
In this figure, $\d =  \frac{1}{2}\e = 1^\circ$.}
\figlab{hexa_cex}
\end{figure}

The key here is the three acute angles of $A$, for it is acute angles
which lead to overlap.
One might be cut (e.g., $a_3$), but that still leaves two to
thwart all placements.
The example can be viewed as an extension of the 
(non-nested) prismatoid described in Figs.~5, 6, and~7 of~\cite{bo-upctt-07},
which has two acute angles, and so thwarted one particular unfolding
but not all unfoldings.


\bibliographystyle{alpha}
\bibliography{/home/orourke/bib/geom/geom}
\end{document}